\documentclass[conference]{IEEEtran}

\usepackage{cite}
\usepackage{amsmath,amssymb,amsfonts}

\usepackage{pgfplots}
\pgfplotsset{compat=1.18}

\usepackage{array}
\usepackage[nolist]{acronym}

\newcounter{savedCounter}

\begin{document}

\begin{acronym}
    \acro{biAWGN}{binary-input additive white Gaussian noise}
    \acro{BPSK}{binary phase shift keying}
    \acro{eBCH}{extended Bose-Chaudhuri-Hocquenghem}
    \acro{GMI}{generalized mutual information}
    \acro{LDPC}{low-density parity-check}
    \acro{MC-DE}{Monte-Carlo density evolution}
    \acro{PPPC}{precoded polar product codes}
    \acro{SC}{successive cancellation}
    \acro{SCL}{successive cancellation list}
    \acro{pm}{path metric}
    \acro{CER}{codeword error rate}
\end{acronym}

\title{Precoded Polar Product Decoder Based on Soft-Output SCL Decoding and Maximization of Generalized Mutual Information}
\author{
	   \IEEEauthorblockN{Nicolás Alvarez Prado,
		                Andreas Straßhofer
		                }
	   \IEEEauthorblockA{Institute for Communications Engineering, 
		                Technical University of Munich,
		                Munich, Germany}
	   \IEEEauthorblockA{\{nicolas.alvarez, andreas.strasshofer\}@tum.de}
}

\maketitle

\begin{abstract}
We combine two approaches to optimize the iterative decoding of product codes with precoded polar component codes. On one side, we generate bitwise soft messages based on the codebook probability, an approximation of an auxiliary quantity that considers all valid decoding paths of a successive cancellation list (SCL) decoder. On the other side, we scale the soft information during message passing with offline-computed coefficients, which maximize the generalized mutual information (GMI) between the channel input and the outgoing message in each half iteration. Simulation results show significant improvement of the error-correcting performance compared to heuristic scaling and soft information generation based solely on the candidate list of the decoder. Moreover, we present an extrinsic version of the SCL decoder, which we use in a Monte Carlo density evolution analysis to derive decoding thresholds. The computed thresholds accurately predict the performance of the decoder.
\end{abstract}

\section{Introduction}
Product codes are a type of error-correcting code that achieve arbitrarily low error probability with strictly positive code rate \cite{Elias54}. They consist of short component codes which iteratively exchange soft information to decode the channel output. Some of the most prominent product codes use \ac{eBCH} \cite{Pyndiah98}, Hamming \cite{Xu07}, or Reed-Solomon \cite{Zhou07} component codes.

Polar codes are the first class of structured codes that provably achieve capacity \cite{Arikan09}. This is made possible through the principle of channel polarization, which synthesizes virtual channels with either very high or very low information rates. The \ac{SC} decoder exploits this structure, with the drawback that it decodes bits in a sequential manner. Since sequential bit decoding is prone to error propagation, polar codes did not gain popularity until the \ac{SCL} decoder was introduced, which circumvented this problem by keeping a memory of candidate codewords \cite{Tal15}. As a consequence of their excellent performance, polar codes were also considered as components for product codes in high rate applications \cite{Ka18, Bioglio19}. However, due to poor distance properties, their performance degrades when considering lower rates \cite{Condo20}.

Product codes with precoded polar component codes overcome this problem \cite{Coskun24}. By introducing dynamic frozen bits into the code design, the precoding step improves the distance properties of the polar component codes \cite{Miloslavskaya20}, hence achieving promising performance even at low rates. 

In this work, we aim to further enhance the decoding of precoded polar product codes. We generate more accurate a-posteriori information by making use of the codebook probability, which is based on approximating the likelihood of each valid path during \ac{SCL} decoding \cite{Yuan23, Yuan24}. Moreover, we use scaling coefficients obtained by maximizing the \ac{GMI}, as suggested in \cite{Strasshofer23}. Our combined method yields gains up to $0.4$ dB in comparison to decoding without consideration of the codebook probability and scaling with heuristically obtained coefficients. We also outperform widely used product codes with \ac{eBCH} component codes. To validate our method we compute decoding thresholds through a \ac{MC-DE}, for which we present an extrinsic version of the \ac{SCL} decoder.

This paper is organized as follows. We cover preliminaries in Section \ref{preliminaries}. We propose the optimized decoder design for precoded polar product codes in Section \ref{iterative_decoding}. The \ac{MC-DE} is presented in Section \ref{decoding_thresholds}, along with the extrinsic version of the \ac{SCL} decoder. We show numerical results in Section \ref{results}.

\section{Preliminaries}\label{preliminaries}
\subsection{Notation and System Model}\label{system_model}
We denote scalar quantities by lowercase letters. Bold lowercase letters represent vectors, for which we use a superscript to specify the length of the vector, e.g., $\boldsymbol{x}^i = (x_1,\dots x_i)$. We omit the superscript in case the vector is of length $N$. Bold capital letters are used for matrices, and their dimensions are to be derived from the context.

We consider transmission over the \ac{biAWGN} channel, i.e., the channel observation is $y = x + z$, where $x \in \{-1, 1\}$ is the transmitted \ac{BPSK} modulated symbol, and $z$ represents the Gaussian noise with zero mean and variance $\sigma^2$. We compute channel log-likelihood ratios (LLRs) as $l^{ch} = \frac{2}{\sigma^2}y$.
\subsection{Product Codes}\label{product}
We focus on two-dimensional product codes. The encoding is carried out as follows. A stream of $k$ bits is rearranged into a $k_2 \times k_1$ matrix. Then, each row of the matrix is encoded with the component code $\mathcal{C}_1$, resulting in a $k_2 \times N_1$ matrix. The product codeword is obtained by encoding all columns using the component code $\mathcal{C}_2$, yielding an $N_2 \times N_1$ matrix.

\subsection{Polar Codes}\label{polar}
Consider a binary memoryless symmetric channel $W$ with transition probabilities 
\begin{equation}
    W(y|x) \triangleq P(Y = y | X = x), \quad \forall (x, y) \in \{-1,+1\} \times \mathcal{Y},  
\end{equation}

and the so-called \emph{polarization kernel}
\begin{equation}
    \boldsymbol{K}_2 \triangleq \begin{bmatrix}
    1 & 0\\
    1 & 1
    \end{bmatrix}.
\end{equation}

The principle of polarization states that applying the $n$-fold Kronecker product of the polarization kernel, i.e., 
\begin{equation}
    \boldsymbol{K}_2^{\otimes n} = \overbrace{ \boldsymbol{K}_2 \otimes \boldsymbol{K}_2 \otimes \dots \otimes \boldsymbol{K}_2}^{n \text{ times}}
\end{equation}

synthesizes $N = 2^n$ virtual channels with different information rates. For large $n$, the reliability values of the synthesized channels tend to either $0$ or $1$, which translates into \emph{useless} or \emph{noiseless} virtual channels, respectively.

Polar codes make use of this property to transmit information bits over the most reliable channels, and \emph{frozen bits}, i.e., bits with fixed values, over the least reliable channels. Polar encoding is thus performed by placing $k$ information bits into the most reliable positions of an $N$-dimensional vector $\boldsymbol{u}$, and subsequently performing the operation $\boldsymbol{c} = \boldsymbol{u}\boldsymbol{K}_2^{\otimes n}$.

\subsection{Precoding Matrices}
Rather than considering only information and frozen bits, one can also make use of \emph{dynamic frozen bits}, where bit values are set to linear combinations of information bits. The mapping from $k$ information bits to a vector which consists of $N$ information, frozen, and dynamic frozen bits, can be carried out with the help of a $k \times N$ precoding matrix $\boldsymbol{P}$. Following the design in \cite{Miloslavskaya20}, such matrices achieve the lowest possible \ac{SC} decoding error probability and guarantee a preselected minimum distance of the resulting code.

The generator matrix of a precoded polar product code can therefore be given as $\boldsymbol{G} = \boldsymbol{P}\boldsymbol{K}_2^{\otimes n}$

\subsection{SCL Decoding}
\ac{SC} decoding \cite{Arikan09} decides on each bit in a sequential manner. The decoder decides for $0$ in case of a frozen bit, takes a linear combination of previously decoded information bits for the case of a dynamic frozen bit, and uses the decision function

\begin{equation}\label{eq:decision function}
    f_i(\boldsymbol{y}, \hat{\boldsymbol{u}}^{i-1}) \triangleq
    \begin{cases} 
        0 & \text{if } P(u_i=0|\boldsymbol{y},\hat{\boldsymbol{u}}^{i-1})  \geq \frac{1}{2}\\
        1 & \text{otherwise},
    \end{cases}
\end{equation}

for the case of an information bit, where the probabilities $P(u_i=0|\boldsymbol{y},\hat{\boldsymbol{u}}^{i-1})$ are computed recursively. The recursive computation of the partial sequence $\tilde{\boldsymbol{u}}^i \in \{0,1\}^i$, is given by
\begin{equation}\label{eq:myop_prob}
    P^{(i)}(\tilde{\boldsymbol{u}}^i \mid \boldsymbol{y}) \triangleq P^{(i-1)}(\tilde{\boldsymbol{u}}^{i-1}\mid \boldsymbol{y})P(\tilde{u}_i\mid \boldsymbol{y}, \tilde{\boldsymbol{u}}^{i-1}),
\end{equation}

where the right-most term is approximated by the \ac{SC} decoder, and an initial probability of $P^{(0)}(\O \mid \boldsymbol{y}) = 1$ is assumed.

\ac{SC} decoding is sub-optimal in the sense that a single mistaken bit decision may propagate over the rest of the algorithm, thus affecting its performance. \ac{SCL} decoding mitigates this drawback by considering a list of decoding paths, each one with an associated path metric $\text{pm}(\boldsymbol{c})$, which, after the last decoding stage, fulfills the relation

\begin{equation}\label{path metric}
    P(\boldsymbol{c} \mid \boldsymbol{y}) = \exp(-\text{pm}(\boldsymbol{c})).
\end{equation}

At each decoding step, if the number of decoding paths exceeds a preset list size $L$, then the decoding paths associated with the highest path metrics are discarded. After decoding the $N$-th bit, \ac{SCL} yields a list of $L$ candidate codewords.

\subsection{Pyndiah-Style Soft-Output Generation}

One way of approximating bit-wise a-posteriori information is based on the output list of the \ac{SCL} decoder as follows

\begin{equation}\label{lapp1}
    L_{i,j}^{\text{app}} \approx \ln \frac{\sum_{\boldsymbol{c} \in \mathcal{L}_{j, 0}^{(i)}} P (\boldsymbol{c} \mid \boldsymbol{l}^{\text{in}})}{\sum_{\boldsymbol{c} \in \mathcal{L}_{j, 1}^{(i)}} P (\boldsymbol{c} \mid \boldsymbol{l}^{\text{in}})},
\end{equation}

where the term $\mathcal{L}_{j,c_j}^{(i)}$ stands for a sublist which includes codewords that, at bit position $j$ have a bit value of $c_j$, and the a-posteriori probabilities associated with each candidate codeword can be expressed as in Eq. \ref{path metric}. 

If one of the sublists is empty, the approximation
\begin{equation}\label{lapp2}
    L_{i, j}^{\text{app}} \approx x_{j} \big( \max_{\boldsymbol{c} \in \mathcal{L}^{(i)}}  \text{pm}(\boldsymbol{c})  - \min_{\boldsymbol{c} \in \mathcal{L}^{(i)}} \text{pm}(\boldsymbol{c}) \big)
\end{equation}
 can be used instead, which weights the sign of the agreed-upon BSPK mdulated codebit with the difference between the maximum and minimum path metrics in the output list \cite{Coskun24}.

\section{Proposed Decoder}\label{iterative_decoding}

\begin{figure*}[!t]
    \setcounter{savedCounter}{\value{equation}}
    \setcounter{equation}{12}
    \begin{equation}
    \label{eq:soscl_app}
    L_{i,j}^{\text{app}} \approx \ln \frac{\sum_{\boldsymbol{c} \in \mathcal{L}_{j, 0}^{(i)}} P (\boldsymbol{c} \mid \boldsymbol{l}^{\text{in}}) + \left(  Q_{\mathcal{U}}^{\star}(\boldsymbol{l}^{\text{in}}) - \sum_{\boldsymbol{c} \in \mathcal{L}_{j}^{(i)}} P (\boldsymbol{c} \mid \boldsymbol{l}^{\text{in}}) \right) \cdot P(c_i=0 \mid l_i^{\text{in}})}{\sum_{\boldsymbol{c} \in \mathcal{L}_{j, 1}^{(i)}} P (\boldsymbol{c} \mid \boldsymbol{l}^{\text{in}}) + \left(  Q_{\mathcal{U}}^{\star}(\boldsymbol{l}^{\text{in}}) - \sum_{\boldsymbol{c} \in \mathcal{L}_{j}^{(i)}} P (\boldsymbol{c} \mid \boldsymbol{l}^{\text{in}}) \right) \cdot P(c_i=1 \mid l_i^{\text{in}})}
    \end{equation}
    \setcounter{equation}{\value{savedCounter}}
    \hrulefill
\end{figure*}
 
Consider the LLR matrix $\boldsymbol{L}^{\text{ch}}$ received after transmitting a product codeword over the biAWGN channel. For each half-iteration step we set the input information to the sum of the channel LLRs and a-priori information from the past half-iteration $\boldsymbol{L}^{\text{in}} = \boldsymbol{L}^{\text{ch}} + \boldsymbol{L}^{\text{a}},$ and for the first half-iteration we set $\boldsymbol{L}^{\text{a}} = \boldsymbol{0}$. Without loss of generality, decoding starts with a row half-iteration, in which we generate bit-wise maximum-a-posteriori information per row, obtaining the a-posteriori matrix $\boldsymbol{L}^{\text{app}}$. The decoded word $\hat{\boldsymbol{C}}$ is obtained by taking hard decisions on $\boldsymbol{L}^{\text{app}}$. If $\hat{\boldsymbol{C}}$ is a valid product codeword, which can be verified by the frozen constraints or the parity-check matrix, then decoding is stopped early. Otherwise, the extrinsic information, computed as
\begin{equation}\label{eq:extrinsic}
    \boldsymbol{L}^{\text{e}} = \boldsymbol{L}^{\text{app}} - \boldsymbol{L}^{\text{ch}} - \boldsymbol{L}^{\text{a}},
\end{equation}

becomes the a-priori information for the following column half iteration

\begin{equation}
    \boldsymbol{L}^{\text{a}} = \alpha_\ell \boldsymbol{L}^{\text{e}},
\end{equation}

where $\alpha_\ell$ is an offline-computed correction factor.

\subsection{Soft-Output SCL Component Decoding}

In an effort to bring the approximations from (\ref{lapp1}) and (\ref{lapp2}) closer to true a-posteriori information, we consider the \emph{codebook probability} \cite{Yuan24}. The codebook probability is an auxiliary quantity given by
\begin{align}\label{eq:cb_prob}
Q_{\mathcal{U}}(\boldsymbol{l}^{\text{in}}) = &\overbrace{\sum_{\boldsymbol{u} \in \mathcal{V}} P\left( \boldsymbol{u} | \boldsymbol{l}^{\text{in}} \right)}^{\text{(a) all visited leaves}} \nonumber \\
&+ \overbrace{\sum_{\boldsymbol{a}^i \in \mathcal{W}} \underbrace{\sum_{\substack{\boldsymbol{u} \in \mathcal{U} \\ \boldsymbol{u}^i = \boldsymbol{a}^i}} P \left( \boldsymbol{u} | \boldsymbol{l}^{\text{in}} \right)}_{\text{(c) all valid leaves underneath node } \boldsymbol{a}^i}}^{\text{(b) all unvisited valid leaves}},
\end{align}

where, in the SCL decoding tree, $\mathcal{V}$ represents the set of visited leaves, $\mathcal{W}$ represents the set of roots of unvisited trees due to complexity limitations, and $\mathcal{U}$ stands for the set of all valid leaves within unvisited trees.

The computation of the codebook probability is possible when assuming that all valid leaves underneath a pruned node $\boldsymbol{a}^i$ are uniformly distributed. Considering that each frozen bit position after the root of an unvisited tree invalidates one half of its leaves, the term (c) in Eq. (\ref{eq:cb_prob}) can be approximated as
\begin{equation}\label{eq:soscl_approx}
    \sum_{\substack{\boldsymbol{u} \in \mathcal{U} \\ \boldsymbol{u}^i = \boldsymbol{a}^i}} P \left( \boldsymbol{u} | \boldsymbol{l}^{\text{in}} \right) \approx 2^{- \left| \mathcal{F}^{(i:N)}  \right|} P(\boldsymbol{u}^i = \boldsymbol{a}^i | \boldsymbol{l}^{\text{in}}),
\end{equation}

where $\left| \mathcal{F}^{(i:N)}  \right|$ denotes the cardinality of the set of frozen bit positions between bit position $i$ and $N$.

\addtocounter{equation}{1}
The approximation for the codebook probability $Q_{\mathcal{U}}^{\star}(\boldsymbol{l}^{\text{in}})$ can be used to generate bitwise soft-outputs as in (\ref{eq:soscl_app}), where the additional term in comparison to (\ref{lapp1}) adjusts the weight between information obtained by the output list from the SCL decoder and the approximated full codebook information. This method of computing a-posteriori information is called soft-output SCL (SO-SCL) decoding.

\begin{figure}
    \centering
    \includegraphics[width=0.9\linewidth]{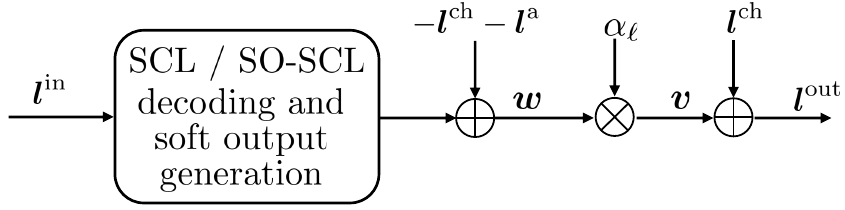}
    \caption{Setup for a decoding half iteration including soft information post processing.}
    \label{fig:decoding_setup}
\end{figure}

\subsection{GMI Based Post Processing}

The extrinsic information matrix from (\ref{eq:extrinsic}) is not a true LLR, since it violates the \emph{consistency condition} \cite{Richardson00}. Also, product code decoding leads to message correlation after a high number of iterations due to cycles in the code's graph. For these reasons, scaling the extrinsic information estimates is widely used to improve performance of product code decoding. In the introduction of precoded polar product codes, the post processing of soft information is based on heuristic coefficients \cite{Coskun24}. We replace the heuristic coefficients with scaling parameters that maximize the GMI \cite{Strasshofer23}.

Figure \ref{fig:decoding_setup} illustrates the flow of a decoding half-iteration, where $\boldsymbol{w}$ and $\boldsymbol{v}$ represent the extrinsic and a-priori information estimates, respectively. We aim to optimize the scaling coefficient $\alpha_\ell$ such that the GMI between the BPSK-modulated input bit and outgoing message in the $\ell$-th decoding half iteration is maximized. We do so by solving the optimization problem
\begin{equation}\label{eq:alpha}
\alpha_\ell^\star = \underset{\alpha_\ell}{\text{arg max}} \left[ 1 - \frac{1}{N_s} \sum_{i=1}^{N_s} \log_2 \left( 1 + \exp\left( -l^\text{out}_i\right)\right) \right],
\end{equation}

where we take into account that the random variable $L^{\text{out}}$ is not a true LLR but an approximation, and therefore we refer to objective function as GMI.

We perform an offline optimization of the scaling coefficients as follows. We generate a large enough number of precoded polar product codewords. For the $\ell$-th half iteration, we generate extrinsic information estimates for each word, and use them to obtain the expression $\boldsymbol{l}^{\text{out}} = \alpha_\ell\boldsymbol{w} + \boldsymbol{l}^{\text{ch}}$, which we plug into Eq. \ref{eq:alpha} to find the optimal alpha.

\begin{figure}
    \centering
    \includegraphics[width=0.89\linewidth]{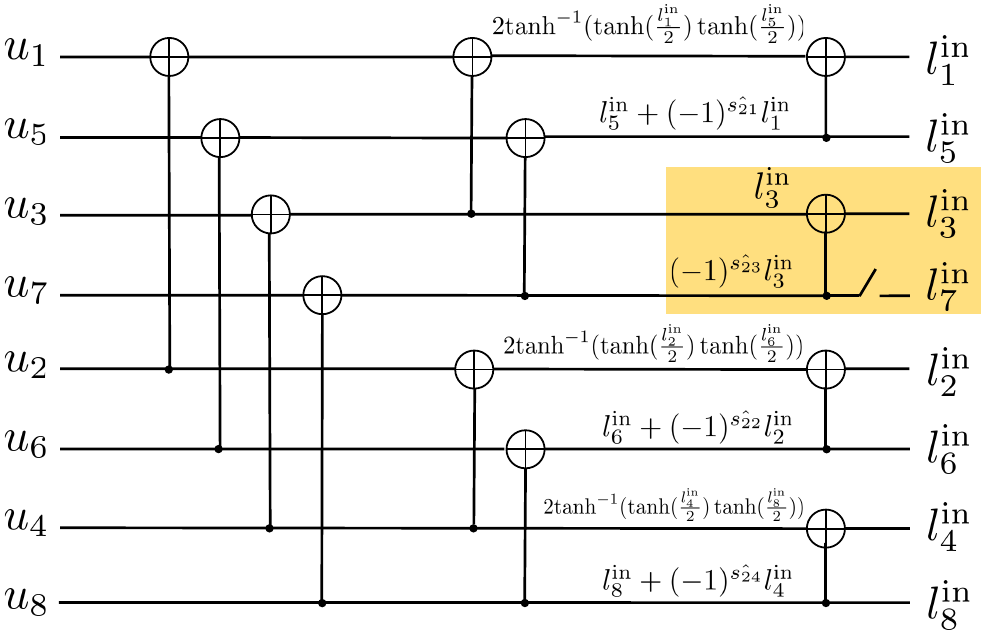}
    \caption{Polar decoding structure under extrinsic SCL for bit position $e=7$.}
    \label{fig:escl}
\end{figure}

\section{Iterative Decoding Thresholds}\label{decoding_thresholds}

\subsection{Monte Carlo Density Evolution}
A density evolution (DE) is performed to compute decoding thresholds, which approximately predict the onset of the waterfall region of the error rate curve. We aim to track the distribution $p_V^{(\ell)}$ of the post-processed output $v = \alpha_\ell w$ throughout the half iterations $\ell \in \{1, 2, \dots, \ell_\text{max}\}$ and under the assumption of transmitting the all-zero codeword. The decoding threshold is defined as the lowest SNR value for which the probability $\text{Pr}[V^{(\ell)} < 0]$ tends to zero for increasing half iterations. Since we do not know the underlying distribution of the post processed messages $V^{(\ell)}$, we use a Monte Carlo approach. We randomly permute the input information to obtain approximately independent incoming messages for the subsequent half iteration.

\begin{figure}[!t]
    \centering
    
    \begin{tikzpicture}
    \begin{semilogyaxis}[
        width=\columnwidth, height=6.5cm,
        xlabel={$E_\text{b}/N_0$ in dB},
        ylabel={CER},
        grid=both,
        legend style={at={(1.03,1)},anchor=north west, font=\footnotesize},
        legend cell align={left},
        ymin=1e-4, ymax=1,
        xmin=1.6, xmax=2.8,
        xtick={ 1.6,  1.8,  2,  2.2,  2.4,  2.6,  2.8},
        ytick={1e-6, 1e-5,1e-4,1e-3,1e-2,1e-1, 1},
        ticklabel style={font=\footnotesize},
        label style={font=\footnotesize},
    ]

    \addplot[color=green!70!black, mark=*, mark options={solid,fill=white}, line width=1] coordinates {
        (1.6, 8.79e-01)
        (1.8, 5e-01)
        (1.9, 3.01e-01)
        (2, 1.09e-01)
        (2.1, 3.9e-02) 
        (2.2, 8.467e-03)
        (2.3, 1.6e-03)
        (2.4, 1.3e-04)
    };\label{line_cp}

    \addplot[color=red, mark=*, mark options={solid,fill=white}, line width=1] coordinates {
        (1.6, 0.69444444)
        (1.8, 0.234)
        (1.9, 0.088)
        (2, 0.02721292)
        (2.1, 0.00559105)
        (2.2, 0.0016875)
        (2.3, 0.00015985)
    };\label{line_gmi_soscl}

    \addplot[color=red, mark=*, mark options={solid,fill=white}, dashed, line width=1] coordinates {
        (1.7, 0.81349206)
        (1.8, 0.60283546)
        (1.9, 0.34914137)
        (2, 0.14329415)
        (2.1, 0.03994722)
        (2.2, 0.00699329)
        (2.3, 0.00103967)
        (2.37, 0.00022996)
    };\label{line_soscl}

    \addplot[color=blue, mark=*, mark options={solid,fill=white}, line width=1] coordinates {
        (1.9, 0.885)
        (2, 0.71825396825)
        (2.1, 0.47619048)
        (2.2, 0.286)
        (2.3, 0.11904762)
        (2.4, 0.04096723)
        (2.5, 0.0105)
        (2.6, 0.0021)
        (2.7, 0.00015995)
        (2.8, 0.0000133)
    };\label{line_heuristic}
    
    \end{semilogyaxis}
    \end{tikzpicture}
    
    \caption{CER performance of ($64^2, 51^2, 6^2$) product codes. Product codes with precoded polar components: heuristic scaling with SCL decoding (\ref{line_heuristic}), heuristic scaling with SO-SCL decoding (\ref{line_soscl}), and GMI scaling with SO-SCL decoding (\ref{line_gmi_soscl}), all considering a list size of $L=8$. A product code consisting of eBCH component codes is also considered, decoded with Chase-Pyndiah decoding with $p=5$ (\ref{line_cp}). We run 20 half iterations.}
    
    \label{fig:6451}
\end{figure}
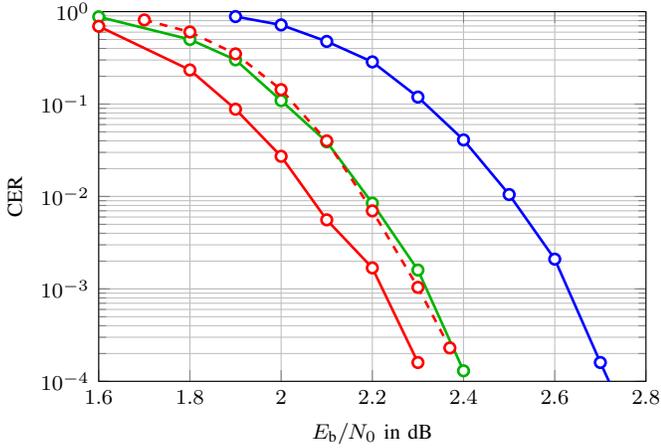

\subsection{Extrinsic SCL Decoding}

DE requires extrinsic message passing between half iterations. The way that soft outputs are generated in (\ref{lapp1}) and (\ref{eq:soscl_app}) does not lead to extrinsic message passing, since SCL and SO-SCL decoding use all bit positions of the incoming message vector $\boldsymbol{l}^{\text{in}}$ to generate the output list.

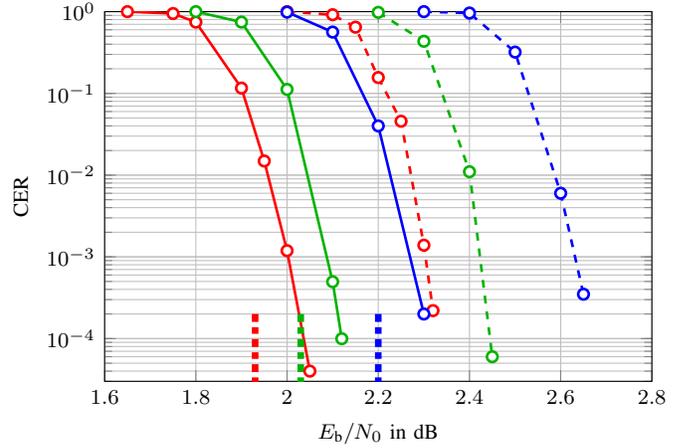
\begin{figure}[!t]
	\centering
	
	\begin{tikzpicture}
		\begin{semilogyaxis}[
			width=\columnwidth, height=6.5cm,
			xlabel={$E_\text{b}/N_0$ in dB},
			ylabel={CER},
			grid=both,
			legend style={at={(1.03,1)},anchor=north west, font=\footnotesize},
			legend cell align={left},
			ymin=3e-5, ymax=1,
			xmin=1.6, xmax=2.8,
			xtick={1.6, 1.8, 2, 2.2, 2.4, 2.6, 2.8, 3, 3.2},
			ytick={1e-6, 1e-5,1e-4,1e-3,1e-2,1e-1, 1},
			ticklabel style={font=\footnotesize},
			label style={font=\footnotesize},
			]
			
			\addplot[color=red, dashed, mark=*, mark options={solid,fill=white}, line width=1] coordinates {
				(2, 1)
				(2.1, 0.916666666666)
				(2.15, 0.6458333)
				(2.2, 0.15625)
				(2.25, 0.045634920635)
				(2.3, 0.0013888888888)
				(2.32, 0.00021993)
			};\label{line_hl8}

			\addplot[color=red, mark=*, mark options={solid,fill=white}, line width=1] coordinates {
				(1.65, 1)
				(1.75, 0.94791667)
				(1.8, 0.74479167)
				(1.9, 0.11607143)
				(1.95, 0.01488095)
				(2, 0.00119048)
				(2.05, 0.00004)
			};\label{line_l8}
			\addplot[line width=2.5, red, no marks, dashdotted] coordinates {(1.93, 2e-4) (1.93, 1e-8)};
			
			\addplot[color=blue, dashed, mark=*, mark options={solid,fill=white}, line width=1] coordinates {
				(2.3, 1)
				(2.4, 0.96153846154)
				(2.5, 0.32)
				(2.6, 0.006)
				(2.65, 0.00034916)
			};\label{line_hl_2}
			
			\addplot[color=blue, mark=*, mark options={solid,fill=white}, line width=1] coordinates {
				(2, 0.98684211)
				(2.1, 0.5625)
				(2.2, 0.04)
				(2.3, 0.0002)
			};\label{line_l2}
			\addplot[line width=2.5, blue, no marks, dashdotted] coordinates {(2.2, 2e-4) (2.2, 1e-8)};
			
			\addplot[color=green!70!black, dashed, mark=*, mark options={solid,fill=white}, line width=1] coordinates {
				(2.2, 0.98)
				(2.3, 0.434)
				(2.4, 0.011)
				(2.45, 0.00006)
			};\label{line_hl4}
			
			\addplot[color=green!70!black, mark=*, mark options={solid,fill=white}, line width=1] coordinates {
				(1.8, 1)
				(1.9, 0.74519231)
				(2.0, 0.11184211)
				(2.1, 0.00049603)
				(2.12, 0.00009968)
			};\label{line_l4}
			\addplot[line width=2.5, green!70!black, no marks, dashdotted] coordinates {(2.03, 2e-4) (2.03, 1e-8)};

		\end{semilogyaxis}
	\end{tikzpicture}

	\caption{CER performance of a ($256^2, 171^2, 12^2$) precoded polar product code decoded using GMI scaling and SO-SCL component decoding with $L=8$ (\ref{line_l8}), $L=4$ (\ref{line_l4}), and $L=2$ (\ref{line_l2}) after $20$ half iterations. Results for SCL component decoding and heuristic scaling (\ref{line_hl8}, \ref{line_hl4}, \ref{line_hl_2}) are provided for comparison. Iterative decoding thresholds are shown as vertical lines.}
	\label{fig:256171}
\end{figure}

We present extrinsic SCL (ESCL) decoding, where we modify the SC decoding algorithm to make sure that the output list used to generate soft information for a specific bit position $e$ is generated by considering all bit positions except for $e$. This is achieved by modifying the $D^-$ and $D^+$ updates steps of the SC decoder at the highest decoding layer, given as
\begin{equation}\label{eq:dminus}
    l_{k-1, j} = 2\tanh^{-1}\left[ \tanh\left(\frac{l_{k,j}}{2}\right)\tanh\left(\frac{l_{k,j+2^{k-1}}}{2}\right)\right],
\end{equation}

and
\begin{equation}\label{eq:dplus}
    l_{k-1, j+2^{k-1}} = (-1)^{\hat{s}_{k-1,j}}l_{k,j} + l_{k,j+2^{k-1}},
\end{equation}

respectively, where the subindex $k$ denotes the SC decoding layer and $j$ is the bit index \cite{Arikan09}.

Consider Eq. \ref{eq:dminus}, as the single-parity check update rule of a degree-$3$ check node, with two input connections and one output connection. We want to neglect the effect of the incoming message value which corresponds to bit position $e$. For this, we simply remove the input connection corresponding to bit position $e$, yielding a degree-$2$ check node with one sole input. Plugging a single input value into Eq. \ref{eq:dminus} results in the input value itself, i.e.,

\begin{equation}\label{eq:new_dminus}
    l_{k-1, j} = 
    \begin{cases}
        l_{k,j} & \text{if $e = j+2^{k-1}$} \\
        l_{k,j+2^{k-1}} & \text{if $e = j$}.
    \end{cases}
\end{equation}

Similarly for Eq. \ref{eq:dplus}, if we remove the connection corresponding to the bit position we want to ignore, we obtain

\begin{equation}
    l_{k-1, j+2^{k-1}} = 
    \begin{cases}
        (-1)^{\hat{s}_{k-1,j}}l_{k,j} & \text{if $e = j+2^{k-1}$} \\
        l_{k,j+2^{k-1}} & \text{if $e = j$}.
    \end{cases}
\end{equation}

An example is provided in Fig. \ref{fig:escl}, which illustrates the SC decoding diagram considering bit reversal permutation. In this case the bit position to be ignored corresponds to $e = 7$, and its connection to the diagram has been removed. It can be observed that, after performing the $D^+$ and $D^-$ updates at the highest decoding layer, the effect of bit position $7$ has been completely neglected. Therefore, the rest of the decoding process can continue as usual.

Note that one run of ESCL allows to compute extrinsic information for a single bit position, unlike classic SCL decoding where with a single run one computes soft information for $N$ bits. For this reason, ESCL decoding is $N$ times more complex, hence unsuited for practical implementation.

\section{Results}\label{results}

\begin{table*}[!t]
    \centering
    \caption{Optimal parameters $\alpha_\ell^\star$ for the $(64^2, 51^2, 6^2)$ precoded polar product code with SO-SCL component decoding, $L = 8$ and $E_b/N_0 = 2.3$ dB throughout half iterations $\ell$.}
    \begin{tabular}{|b{0.3cm}|b{0.4cm}|b{0.4cm}|b{0.4cm}|b{0.4cm}|b{0.4cm}|b{0.4cm}|b{0.4cm}|b{0.4cm}|b{0.4cm}|b{0.4cm}|b{0.4cm}|b{0.4cm}|b{0.4cm}|b{0.4cm}|b{0.4cm}|b{0.4cm}|b{0.4cm}|b{0.4cm}|b{0.4cm}|b{0.4cm}|}
        \hline
        $\ell$ & 1 & 2 & 3 & 4 & 5 & 6 & 7 & 8 & 9 & 10 & 11 & 12 & 13 & 14 & 15 & 16 & 17 & 18 & 19 & 20 \\ \hline
        $\alpha_\ell^\star$ & 0.96 & 0.92 & 0.88 & 0.83 & 0.75 & 0.69 & 0.66 & 0.60 & 0.59 & 0.58 & 0.60 & 0.51 & 0.52 & 0.48 & 0.47 & 0.46 & 0.51 & 0.49 & 0.51 & 0.47 \\ \hline
    \end{tabular}
    \label{tab:soscl_params}
    \vspace{5mm}
    \hrule
\end{table*}

Figure \ref{fig:6451} shows the \ac{CER} decoder performance of $(64^2, 51^2, 6^2)$ product codes, which correspond to a code rate of $0.635$. We consider precoded polar product codes with our optimized decoder, i.e., with SO-SCL component decoding and GMI-based scaling coefficients, and compare them against the decoder presented in \cite{Coskun24}, which uses SCL component decoding and heuristic scaling. The precoding matrices are optimized for SC decoding and obtained from \cite{Miloslavskaya20}. We also compare our results against product codes with \ac{eBCH} component codes, which employ Chase-Pyndiah decoding with the number of least reliable bit positions set to $p=5$. Our approach yields gains of up to $0.4$ dB in comparison to the decoder without our optimizations. Moreover, we outperform product codes with \ac{eBCH} components by around $0.1$ dB. Note that for the product codes with precoded polar component codes, the \ac{SCL} decoder takes into account $L=8$ candidate codewords, whereas for Chase-Pyndiah decoding the Chase decoder considers up to $2^p = 32$ codewords.

In Table \ref{tab:soscl_params} we provide the computed scaling coefficients that maximize the GMI throughout the half iterations for the $(64^2, 51^2, 6^2)$ product code with precoded polar component codes at $E_\text{b}/N_0 = 2.3$ dB and $L=8$. A remarkable observation is that the coefficients start with values close to one, and the values decrease with increasing half iterations. We recall that the use of SO-SCL decoding aims to bring the generated soft outputs closer to true a-posteriori information by considering the codebook probability. This effect might be an explanation to the values close to one for the first few half iterations. Recall that cycles in the product code's graph lead to message correlation after some half iterations, which might explain the decreasing values of the scaling coefficients.

\ac{CER} curves for a $(256^2, 171^2, 12^2)$ precoded polar product code with our optimized decoder and different list sizes are presented in Fig. \ref{fig:256171}. The code rate is $0.446$. We show the performance of the decoder without our optimizations in dashed curves for comparison. For list sizes of $L=8$ and $L=4$ the gains with respect to the dashed curves are around $0.3$ dB, while for a list size of $L=2$ we observe a gain of almost $0.4$ dB. The solid blue error curve, corresponding to $L=2$ with our decoder optimizations, outperforms the decoder with a list size of $L=8$ without our optimizations, i.e., \ac{SCL} component decoding and heuristic scaling. This means that our methods allow for a complexity reduction of factor $4$ with respect to the legacy precoded polar product code decoder, without comprising performance. We illustrate as vertical lines the decoding thresholds obtained through MC-DE with ESCL decoding, for which we use $10^5$ samples and run $50$ half iterations. It can be observed that the decoding thresholds not only accurately predict the onset of the waterfall region of the error curves, but also manage to predict the performance of the product code, in the sense that the gains between the \ac{CER} curves correspond to the distance between thresholds. We remark that in order to compute the iterative decoding thresholds using MC-DE analysis we use the product code rate rather than the design rate of the underlying code ensemble.

\section{Conclusion}\label{conclusion}
We combined two approaches to the decoding of precoded polar product codes, namely the use of the codebook probability for more accurate generation of a-posteriori information and the optimization of the scaling coefficients based on the GMI. The proposed optimizations yield significant gains compared to existing decoders. We introduced an extrinsic version of the SCL decoder, which allows performing an MC-DE analysis to find iterative decoding thresholds.

\end{document}